\documentclass[sigconf]{acmart}

\AtBeginDocument{%
  \providecommand\BibTeX{{%
    \normalfont B\kern-0.5em{\scshape i\kern-0.25em b}\kern-0.8em\TeX}}}

\copyrightyear{2019}
\acmYear{2019}
\setcopyright{acmlicensed}
\copyrightyear{2019} 
\acmYear{2019} 
\acmConference[GoodTechs '19]{EAI International Conference on Smart Objects and Technologies for Social Good}{September 25--27, 2019}{Valencia, Spain}
\acmBooktitle{EAI International Conference on Smart Objects and Technologies for Social Good (GoodTechs '19), September 25--27, 2019, Valencia, Spain}
\acmPrice{15.00}
\acmDOI{10.1145/3342428.3342683}
\acmISBN{978-1-4503-6261-0/19/09}

\usepackage{booktabs}
\usepackage{amsmath}
\usepackage[]{algorithm2e}

\begin{document}

%
\title[A Musical Serious Game]{A Musical Serious Game for Social Interaction through Augmented Rhythmic Improvisation}

\newcommand{\myuni}{University of Padova (IT)}
\newcommand{\mydept}{Dep. of Information Engineering}
\newcommand{\mycity}{Padova (IT)}
\newcommand{\mycountry}{Italy}
%
\author{Filippo Carnovalini}
\email{filippo.carnovalini@dei.unipd.it}
\orcid{0000-0002-2996-9486}
\affiliation{%
  \institution{\mydept}
  \institution{\myuni}
}

\author{Antonio Rod\`a}
\email{roda@dei.unipd.it}
\affiliation{%
  \institution{\mydept}
  \institution{\myuni}
}

\author{Paolo Caneva}
\email{paolo.caneva@conservatorioverona.it}
\affiliation{%
  \institution{Department of Music Therapy}
  \institution{Verona Conservatory of Music (IT)}
}

%

%
\begin{abstract}
Making music with other people is a social activity as well as an artistic one. Music therapists take advantage of the social aspects of music to obtain benefits for the patients, interacting with them musically, but this activity requires an high level of expertise. We propose a serious game that helps people even without musical skills interact with each other by collaboratively creating a rhythm with MIDI drum pads. The gaming system analyzes the rhythm in real time to add a musical feedback that enhances the aesthetical experience that is a crucial part of the musical interaction. The system is evaluated through a questionnaire asking subjects who tried in couples the system if they perceived it as helping their interaction. Despite the early development stage of the game, the results of the questionnaire show positive reception.
\end{abstract}

%
%

%

%

%
\maketitle

\section{Introduction}
\label{sec:introduction}
A key factor in music playing is the interaction between the musicians: when there is a written score, it is necessary for them to keep the same tempo in order to effectively play the written music. In the case of improvisation, the interaction becomes even more evident: even if a common ground is chosen beforehand (like a fixed chord progression for example) it is necessary for the musicians to proactively listen to what the others are playing to obtain good musical results. 
This interactivity of music playing is the basis of many of its recognized educational benefits: the ``social'' aspect of music develops a sense of belonging and strengthens the social skills as well as self-esteem and satisfaction~\cite{kokotsaki_higher_2007,hallam_power_2010}.    
The emotional value of music is also a key factor in the emergence of these social benefits~\cite{koelsch_neuroscientific_2009,koelsch_music-evoked_2015} and can be helpful even with subjects with social difficulties, for example those suffering from autism spectrum disorders (ASD), that can express feelings through music more effectively than with words~\cite{allen_autism_2010,heaton_can_1999,quintin_emotion_2011}. 
These benefits are widely recognized in young musicians playing in orchestras or bands, but the training needed to obtain an effective interaction and good musical results can be very demanding. Such an high level of technical ability becomes impossible to achieve for people with motor impairments, or simply for those that do not wish to spend lot of time learning how to play a musical instrument.

Professional music therapists can help their clients obtain these kinds of benefits despite the lack of training. There is a great variety of different techniques and methods in music therapy: one is letting the client play a simple instrument, for example a drum, that does not require high musical skills, while the therapist interacts with the patient using more complex instruments, such as a piano, creating music that follows the client's inputs but is far more elaborated and aesthetically pleasing. 
This of course requires the help of a highly trained professional, and is therefore hard to use as a everyday and inexpensive activity. Moreover, this kind of sessions usually consist of a child interacting with the therapist: there could be an interaction with more then one children, but it is impossible to have a ``peer to peer'' dialogue between children alone. 
To obtain an ``expertless'' experience, one could limit the quality of the music produced. For example, if all the participants use only drums, that are less demanding in terms of musical training, one would obtain a rhythmic improvisation rather than a melodic one. This could be a good compromise since the interactivity is maintained, that is seemingly one of the most important feature of the music therapy. Yet, the aesthetic value of the interaction is a crucial factor for the therapeutic effects of a musical dialogue~\cite{aigen_defense_2007, stige_aesthetic_1998}, and should therefore not be overlooked.

In this paper we propose a serious game that has the aim of helping the players obtain a musical interaction without the need for particular expertise. We do not focus on a specific target, as anybody could ideally benefit and enjoy these kinds of interactions, 
but it could be especially fit for children with social difficulties.
The input of the game consists of two MIDI pads, so that the users can only interact in a rhythmic way, as if they were using two drums. The difference from a normal rhythmic improvisation with drums is that this system can `augment' the improvisation with melodic and harmonic aspects, to obtain a more satisfying musical output. 
The goal of the game is the interaction itself: the two players start playing the pads without any musical augmentation, only hearing the sound they make with the percussion. The system then evaluates the quality of the interaction between the two players, and assigns a score to the interaction. This score is presented in two ways: one is the usual visual feedback expressed in ``points'' that are collected by keeping a good interaction over time, and the other is a ``musical'' feedback. This feedback is the musical augmentation: it follows the tempo and rhythm given by the two players, but adds a harmonic and melodic dimension to the musical production. 
The aesthetic value of this output is meant to give a more entraining experience to users, giving an incentive for a better interaction between the players.

There are other works in literature that focus on following a human improvisation, but to our knowledge all those works need some level of prior information to work properly, like a fixed tempo~\cite{hawryshkewich_beatback:_2010}, a predefined chord structure~\cite{dannenberg_following_1987}, or a general melody upon which the improvisation is developed~\cite{xia_improvised_2017}. The system we propose has instead no such need, making it as accessible as possible to people without any musical knowledge. 

\begin{figure}
    \centering
    \includegraphics[width=0.35\textwidth]{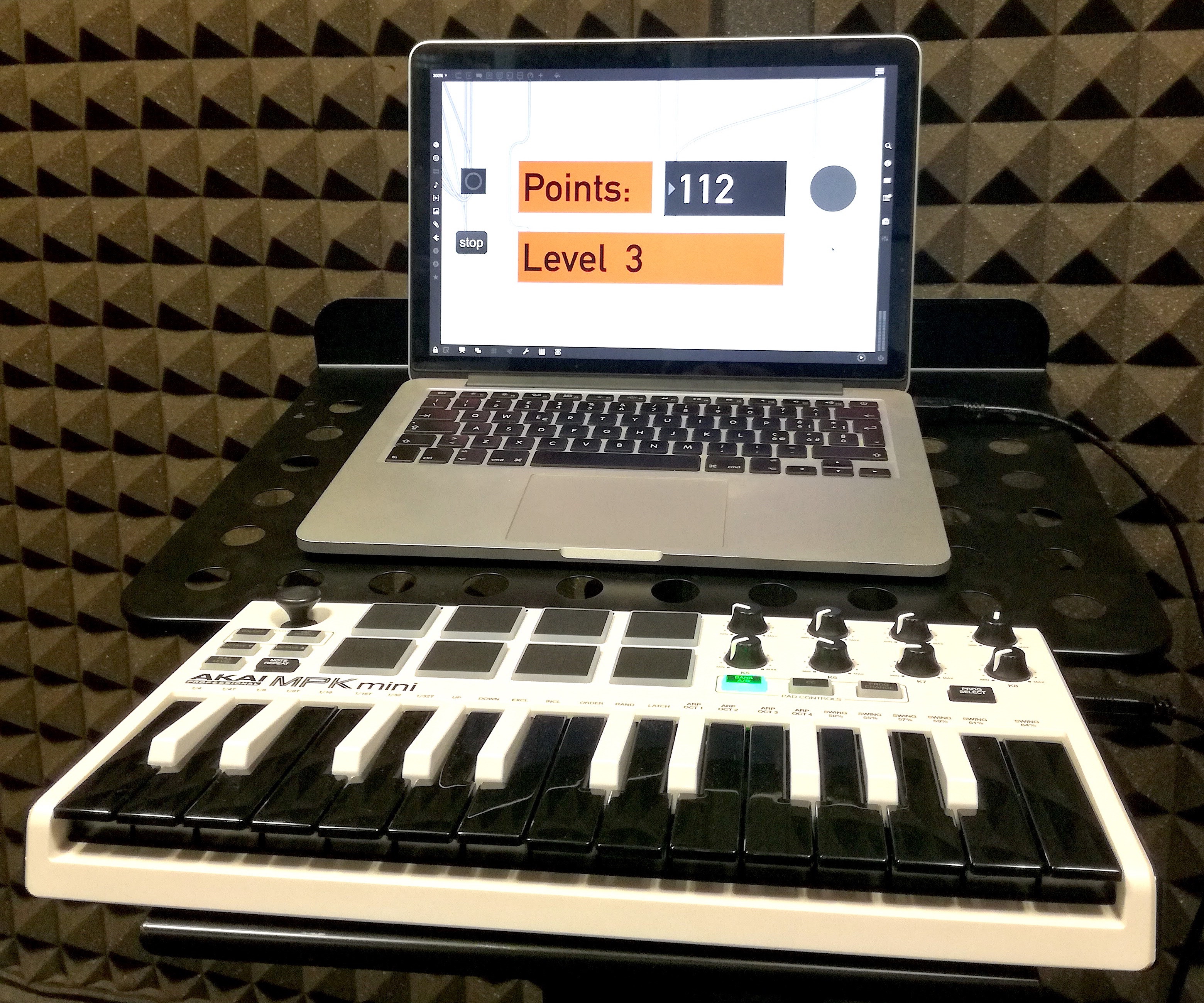}
    \vspace{-0.2cm}
    \caption{The interface of the system: a computer running the software connected via USB to a MIDI keyboard with drum pads.}
    \label{fig:setup}
    \vspace{-8mm}
\end{figure}

\section{Architecture}
\label{sec:architecture}

The system's architecture is divided in three main parts. The first is the ``Listener'' module, that has the aim of collecting the input from the user, and to infer low level features like tempo and measure duration. The second part, the ``Scorer'' module uses this information to compute higher level features that are needed to evaluate the interaction between the two players. The last module, the ``Generator'', uses the information from the other two parts to create a musical output, and to visually notify the users of the quality of their interaction with a number of points. The following three subsections focus each on one of these modules, and the fourth will add some implementation remarks on the whole system. 

\subsection{Listener}
\label{sub:listener}

\begin{figure}[t]
    \centering
    \includegraphics[trim={4.3cm 2.8cm 4cm 2cm}, clip,width=0.4\textwidth]{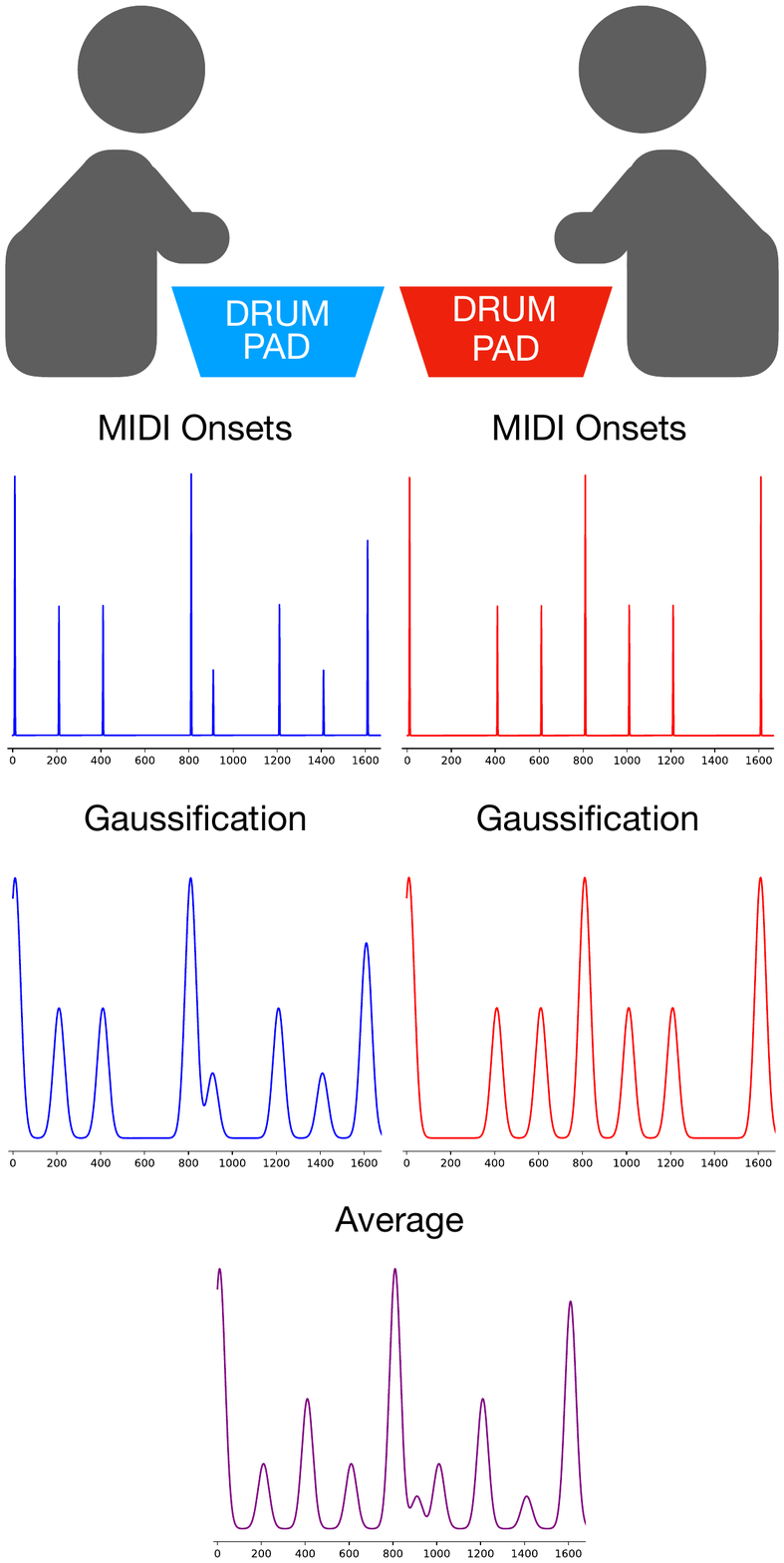}
    \vspace{-0.2cm}
    \caption{Data flow in the Listener module: the onsets and velocities are collected from the drum pads and then gaussified, and an average of the two signals is computed.}
    \label{fig:pads}
    \vspace{-6mm}
\end{figure}

The only input interface of the system is represented by the MIDI pads (the black squares in Figure \ref{fig:setup}) used by the users. Each time a user hits the pad, a timestamp is saved when the note\_on message of the percussion is received, along with the force with which the pad was struck (MIDI velocity). For each pad a list of events is saved, representing the rhythm played by the user: the topmost plots in Figure \ref{fig:pads} show how this input data can be represented. 

The users are not forced to follow any pre-established tempo and do not follow any written indication on what to play. It is thus necessary for the system to infer this information, in order to be able to follow what the users play. To obtain this, the system follows the algorithm by Freiler~\cite{frieler_beat_2004}, that only uses onsets and velocities to infer the quantization of a rhythmic signal, without the need of any other information.

The main idea of this algorithm is that of \textit{gaussification}, that is the construction of an integrable function from a list of timing points by taking the linear combination of the gaussians centered on the input points, as can be seen in the lower plots in Figure \ref{fig:pads}. 
Once this function is constructed, it is possible to infer the beat of the rhythm by computing the autocorrelation of the function at different time points: the highest peak of correlation is the perceived beat. The algorithm also uses a normalization based on perceptive features to ensure that the most appropriate tempo is chosen when there are ambiguities.
The meter (the number of beats in each measure) is computed by comparing ``prototypical'' functions representing different meters with the input signal. While doing this, the algorithm also determines the ``phase'' of the signal, that is the timing of the beginning of the measure, that does not necessarily coincide with the beginning of the signal.

The algorithm uses a single set of time points to compute these features, but our system has two distinct lists of timestamps: one per each user. It would be possible to compute beat and meter for both signals, but users hear sounds coming from both pads and the beat and meter they perceive is based on both the signals. Thus, the average of the two signals is computed, representing the rhythm perceived by listening to both the pads, and beat and meter are computed on this averaged signal. 

The output of this module is the estimate for the tempo and the meter of the average signal, as well a timestamp representing the estimate time of the beginning of the next measure, based on the the estimated start of the measures in the signal (the phase) and the estimated duration of a measure. This prediction is especially important since it is used by the Generator module to synchronize the generated music with the sounds produced by the users playing the pads.

\vspace{-0.2cm}
\subsection{Scorer}
\label{sub:scorer}

The features computed by the Listener module are needed for the correct generation of music and to ensure the synchronization of the generated music with the rhythm played by the users. The features that are considered by the Scorer module are instead more related to the gaming aspect of the system, and are needed to give a feedback to the users about the quality of their interaction. 
The theoretical basis for this module is taken from music therapy, and in particular from the ``Improvisation Techniques for Music Therapy'' devised by Kenneth Bruscia~\cite{bruscia_improvisational_1987}. This is a wide set of possibilities available to the music therapist to obtain a better interaction with the client. These are divided in different categories, and don't focus solely on the music production but also on the physical/visual interaction with the client, that is a kind of information that is not available to our system. We selected five of the main basic features that only rely on the rhythm produced by the players. Here we report the definitions given by Perret in his comment on Bruscia's work~\cite{perret_roots_2005} for the chosen techniques:

\begin{description}
\item[Imitation] Echoing or reproducing a client's response, after the response has been completed;
\item[Synchronization] Doing what the client is doing at the same time;
\item[Incorporating] Using a musical motif or behaviour of the client as a theme for one's own improvising or composing, and elaborating it;
\item[Pacing] Matching the client's energy level (i.e., intensity and speed);
\item[Rhythmic grounding] Keeping a basic beat or providing a rhythmic foundation for the client's improvisation.
\end{description}

These techniques are meant to be used by the music therapist to help the client during the improvisation, and are not immediately fit for our situation where two peers are playing together and must be evaluated by a computer system. Nonetheless, those were useful in designing more precise and measurable features that could be used in our system. In particular, the system distinguishes four possible levels, describing the quality of the interaction:  
\begin{description}
\item[Level 0:] The system is incapable of clearly following the users, as they are not making a clear enough beat (no Synchronization or Rhythmic grounding);
\item[Level 1:] The system is capable of following the users, but one is dominating the rhythm and the other is not contributing (no Pacing);
\item[Level 2:] The interaction is considered normal;
\item[Level 3:] The interaction also includes imitations between the two players (Imitation and Incorporation).
\end{description}

To get a quick idea of the differences between these levels, please visit the video linked in the footnote\footnote{\url{https://mediaspace.unipd.it/media/0_g90zoo2n}}.
The algorithm computes the level of interaction according to Algorithm \ref{alg:levels}, that requires in input the gaussified signals (the ones of the two players as well as the averaged one), the duration of a beat and of a measure in milliseconds, and the list of the timestamps of the notes produced by the two players. The algorithm uses three functions: \textit{correlation(a,b)}, that computes the correlation of the signal \textit{a} with the signal \textit{b}; \textit{now()}, that returns the current timestamp; and \textit{shift(a,b)} that moves along the x axis all the points of the signal \textit{a} by \textit{b} units.

\begin{algorithm}[h]

\SetAlgoLined
\KwIn{$signal_1, signal_2, signal_a, beat, measure, notes_1, notes_2$}
\KwOut{Level of interaction}
$clarity \gets$ correlation($signal_a$, shift($signal_a,beat$))/ correlation($signal_a,signal_a$)\;
 \If{clarity < 0.4}{
    \Return{0}\;
 }
 $density_1 \gets$ |el : el > now()-10000 $\And\And$ el $\in notes_1$|/10\;
 $density_2 \gets$ |el : el > now()-10000 $\And\And$ el $\in notes_2$|/10\;
 \If{$density_1$ < 0.5 || $density_2$ < 0.5}{
    \Return{1}\;
 }
 $crossCorr \gets$ (correlation($signal_1$, shift($signal_2$,measure)) + correlation($signal_2$, shift($signal_1$,measure))) /2\;
 \eIf{crossCorr < 15}{
    \Return{2}\;
 }{
    \Return{3}\;
 }
 \caption{The algorithm for the computation of the current interaction level}\label{alg:levels}
\end{algorithm}

The algorithm computes three significant features to distinguish the levels. The $clarity$ represents how confident the system is in estimating the current beat. The two $density$ values represent the notes per second each user plays. Finally, $crossCorr$ is how similar the two signals are at the distance of one measure, i.e. how much the users are imitating each other measure by measure. 
The threshold values for the various levels were chosen empirically by computing the average values obtained by a ``metronome'' interaction, i.e. where the two users were substituted by a software sending a beat at regular intervals. It would be possible to use machine learning to determine better thresholds if enough data is collected from real users' interactions. 

The final score is given to the users by adding each second a number of points depending on the current level. Notice that since the goal of the game is the interaction between the two players rather than a challenge, the score is shared. 
The values of these features are not directly transformed into points given to the user. The features are use to compute a multiplication level, and a number of points depending on the current level is added to the users' score every second. 
To make sure that scoring system is stable, the level shown in the interface is not the latest computed level, but the median of the fifteen last computed levels. As well as being necessary for the visual feedback, the levels also influence the musical output: the music generation method is the same at every level (as described in the next section), but the volume of the three generated instruments is proportional to the level reached by the players.  

\vspace{-0.2cm}
\subsection{Generator}
\label{sub:generator}

The musical output could potentially be of any nature: the Listener module gives all the necessary information for the synchronization of music to the beat produced by the users, so it would be possible to chose any musical audio file and synchronize it via time warping~\cite{dannenberg_-line_1984,robertson_b-keeper:_2007}. 
In the system we created we decided to use procedurally generated music, and to keep the music simple, in order not to take the focus away from the interaction between the users. If the music becomes too compelling, the users could just follow the music and become a sort of `human metronome', that is not the goal of this system. 
The Generator receives as input all the results of the other modules' computations. The most important information is the prevision of the beginning of the next measure. The Generator saves all the previsions obtained from the Listener, and each time a pad is struck the current time is compared with the previsions: if a prevision is found that is within 50 millisecond from the current time, the system understands that the received input is the beginning of a measure. Every time this happens, the tempo and the meter of the Generator are updated to match the ones computed by the Listener, and the internal metronome of the Generator is reset and started. 
This approach was chosen over the possibility of having the Generator start a measure at the foreseen moment without waiting for an input from the user because this increases the feeling of control over the output. Having the system react to specific actions performed by the user is important to obtain the feeling of ``I made this happen'', that is considered crucial for the effectiveness of music therapy~\cite{swingler_invisible_1998}.

\begin{figure}[t]
    \centering
    \includegraphics[trim={0.8cm 0.5cm 0.4cm 0.3cm}, clip, width=0.25\textwidth]{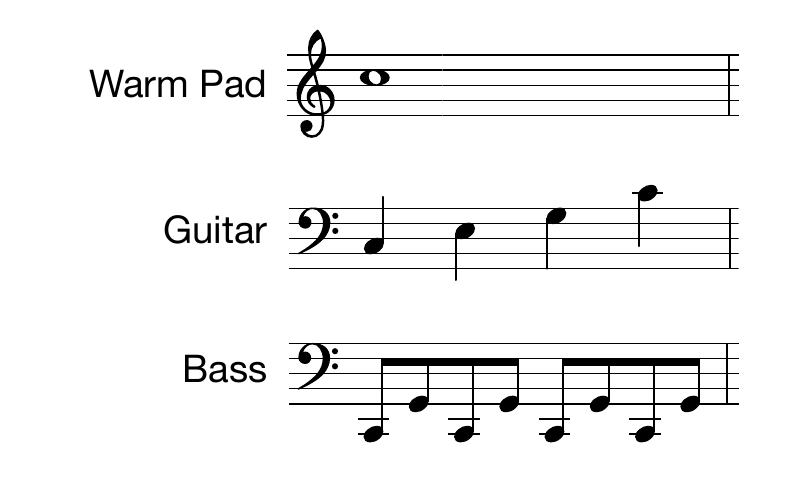}
    \vspace{-0.2cm}
    \caption{A generated measure in 4/4 for each of the instruments, based on a C major chord.}
    \label{fig:contours}
    \vspace{-5mm}
\end{figure}

At the beginning of each measure a chord is selected using a very simple Markov chain, based on Stephen Mugglin's chord progression charts~\cite{stephen_mugglin_music_2017}. The transition matrix is shown in Table \ref{tab:chords}. This generates chords progressions in the tonality of C major, but it would be easy to transpose the selected chords to other tonalities. 

\begin{table}
\begin{tabular}{@{}r|llllll|@{}}
\cmidrule(l){2-7}
\multicolumn{1}{l|}{}    & \multicolumn{1}{c}{C} & \multicolumn{1}{c}{Dm} & \multicolumn{1}{c}{Em} & \multicolumn{1}{c}{F} & \multicolumn{1}{c}{G} & \multicolumn{1}{c|}{Am} \\ \midrule
\multicolumn{1}{|r|}{C}  & 1/6                   & 1/6                    & 1/6                    & 1/6                   & 1/6                   & 1/6                     \\
\multicolumn{1}{|r|}{Dm} & 1/3                   & 0                      & 1/3                    & 0                     & 1/3                   & 0                       \\
\multicolumn{1}{|r|}{Em} & 1/3                   & 0                      & 0                      & 1/3                   & 0                     & 1/3                     \\
\multicolumn{1}{|r|}{F}  & 1/3                   & 1/3                    & 0                      & 0                     & 1/3                   & 0                       \\
\multicolumn{1}{|r|}{G}  & 1/3                   & 0                      & 1/3                    & 0                     & 0                     & 1/3                     \\
\multicolumn{1}{|r|}{Am} & 0                     & 1/2                    & 0                      & 1/2                   & 0                     & 0                       \\ \bottomrule
\end{tabular}

    \caption{The transition matrix of the Markov chain used to create chord progressions in C major.}
    \label{tab:chords}
    \vspace{-8mm}
\end{table}

Depending on the selected chord, three accompaniment instruments are played: a warm pad (General MIDI instrument 90) that plays the tonic, a guitar that plays one note of the chord's arpeggio each beat, and a bass that plays alternating the tonic and the fifth of the chord every quaver. An example of a generated measure for the C major chord in 4/4 time is shown in Figure \ref{fig:contours}.
The chord is changed again each time there is a synchronization with the foreseen beginning of a measure or when the internal metronome of the Generator reaches the 1st beat again (i.e. there is a chord change each measure).

\begin{table*}
\centering
\begin{tabular}{@{}lll@{}}
\toprule
Question                                                                           & Avg. Response & Std. Dev. \\ \midrule
I found the game experience pleasant.                                              & 5.92          & 1.00      \\
I think the music was aesthetically pleasing.                                      & 5.04          & 1.17      \\
I wished to keep playing.                                                          & 5.58          & 1.22      \\
I wish to play again with this system in the future.                               & 5.25          & 1.20      \\
I was actively interacting with the other participant.                             & 5.33          & 1.72      \\
I was actively interacting with the system.                                        & 5.08          & 1.32      \\
The system was interacting actively with me and the other participant.             & 5.50          & 1.29      \\
I felt I had control over what was happening musically.                            & 4.17          & 1.34      \\
The system reacted to what me and the other participant were playing.              & 5.25          & 1.27      \\
The music produced by the system was a stimulus to interact with the other player. & 5.58          & 1.32      \\ 
\bottomrule
\end{tabular}
\caption{The questions posed to the participants of the evaluation of the system with their average response and standard deviation in a scale from 1 (Completely disagree) to 7 (Completely agree).}
    \label{tab:questions}
    \vspace{-8mm}
\end{table*}

\subsection{Implementation}
\label{sub:implementation}

The modules described above represent the abstract functionalities of the system rather than actual software modules. The real implementation we used for this system uses a Max/MSP\footnote{\url{https://cycling74.com/products/max/}} patch for the collection of the input from the user and the generation of the music. The patch communicates via Open SoundControl~\cite{wright_open_1997} with a Python script, that receives the list of inputs from the patch and computes in a stateless function both the features described in the Listener module and those related to the Scorer module. Practically, the Generator is implemented in Max/MSP and the Scorer in Python, but the Listener is shared between the two systems.
The Max/MSP patch also implements the GUI that is shown in Figure \ref{fig:setup}.
This subdivision was chosen because Max/MSP is not fit for computations like those needed for the autocorrelation, but on the other hand Python is not ideal for real-time computing. Having a stateless server invoked by the patch ensures that the computations done by Python are not critically dependent on timing, as all the time labelling is handled by Max/MSP, while keeping the advantage of using Python and NumPy for the computation of the correlation of signals. 
\vspace{-2mm}
\section{Evaluation}
\label{sec:evaluation}

The main goal of this study was to create a serious game that would help the players develop a rhythmic improvisation, entraining and helping the players establish a common ground with the music added by the system. 
In order to test how well the system fulfills these goals, a questionnaire was used as a first form of evaluation of the system. 

Twenty-four participants were collected among university students. Nine were females and fifteen males with the average age of 24.17 years (standard deviation 3.48 years). In couples, the participants were instructed to use the system to create a ``rhythmic interaction''. They were told that the system would add music to their performance based on how they interacted. After playing with the system for three minutes, each of the participants was asked to fill a questionnaire consisting of ten 7-points Likert items. The questionnaire also asked what was their level of musical expertise, but except for one classically trained musician all reported little amateur experience or no experience at all. All the questions were posed in Italian, as it was the native language of all the participants.

The questions in English, as well as the results of the questionnaire, are reported in Table \ref{tab:questions}. The Cronbach's alpha of the collected data is 0.83, showing that the questionnaire has a reasonable reliability. The first four questions were meant to assess if the participants liked to play with the system, while the other questions were more intended to assess if they felt the system helped them interact musically.

On average, the participants found the system to be pleasing to play with and would have liked to play more, meaning that the system results to be entraining. 
The results regarding the interactivity of the system are also positive, showing that the participants felt there was an interaction both between the players and with the system, and that the produced music was helpful to their interaction. Despite the perceived sense that the system reacted to their inputs, the participants did not really feel like having control over the produced music. This is reasonable, as the rhythmic dimension of the music, the one that is directly controlled by the users, is only a fraction of the whole musical output. 

Aside from the questionnaire, more qualitative observations collected during the experiment show that while the system is both capable of following an established rhythm and quickly adapting to changes in tempo, the users are not really driven to experimenting with more complex interactions. This is probably due to the fact that the system takes a few seconds to adapt to unexpected changes, and the immediate feedback (before the system adapts) is negative. This induces the players to stick to the first common rhythm they can establish, that is not necessarily the best situation. 
Moreover, it was noticed that while the system is capable of handling meters different from 4/4, all the participants settled on this meter.
In general, the results collected from the questionnaire are not very strongly detached from the neutral response, showing that there is still much room for improvement in the system. Nonetheless, considered that the evaluation was carried on what is the very first version of the system, the results are very encouraging. Further developments could improve the interaction to give a better feeling of control over the produced music, for example by adding more complex musical variations to the output mimicking the users' input more closely. 

\vspace{-2mm}
\section{Conclusions}
\label{sec:conclusions}

In this paper we described a musical serious game that makes it possible to have interpersonal interactions that are common in music making, even without having special musical expertise or training, and without the intervention of a music therapist. 
The game uses the rhythmic interaction between two players, collected through MIDI drum pads, to determine features that help the system add a musical background to the interaction, with both the goal of rewarding good interactions between the players and of entraining the players.
We described how the architecture of the system works, and how it infers the features it needs by computing the  correlation of signals constructed from the timing and velocities of the players' beats. From the computed correlation it is possible to infer the beat kept by the players as well as the musical meter they are producing, and it is possible to predict the timing of the beginning of the next measure. We also described some metrics extracted from the same signal were used to evaluate the interaction between the two players.

The game was evaluated through a questionnaire: eighteen students played with the system and responded to questions about their experience. The collected answers show that the game was generally enjoyed by the players, and that they felt it actually helped them interact with each other. 

This game was designed having in mind the benefits that musical interactions can have on social skills, especially in children with social difficulties, but the system was designed not to address a specific category. It is a tool that can be used by anyone to musically interact with someone else. 
Other categories of users that could benefit from this kind of musical game include patients recovering from conditions that can impair motor skills, like stroke. The recovery therapies include many repetitive exercises, that can be made less fatiguing with the help of the musical augmentation~\cite{chen_music-supported_2018,fujioka_effects_2018}.
This system could also be used to teach music students how to keep a steady beat~\cite{schaefer_motor_2015}. Obviously, this system can also be used by music therapists to enrich their sessions.

\vspace{-0.3cm}
\subsection{Future Work}

The game presented in this paper still needs more refinements before it can be considered a finished product. Yet, before going further with the development, one could consider evaluating the system in ways complimentary to the one presented in this paper: first of all, the effectiveness of the tempo and meter detection algorithm wasn't directly evaluated, if not through informal testing. Also, what we collected were subjective impressions of users, while psychophysiological data could be more insightful and reliable indicators of entrainment and emotional activation. Also, once a specific category of users is fixed and goals concerning and the desired therapeutic effects of the game are established, a long term evaluation is needed to assess if the benefits that usually come from music therapy can be obtained with this system. 

As pointed out in Section \ref{sec:evaluation}, the game would benefit from more complex variations in the generated music, to be more capable of mimicking the users' inputs. This requires advancements both in the Scorer module, that would need to extract more and more meaningful features, and in the Generator module, that should be capable of adapting the generated music to those features. For example, the articulation and the velocity of the produced notes could vary depending on the force used by the users, as these are two clear indicators of emotional intent~\cite{juslin_expression_2010,sab12,canazza_caro_2015,Turchet:2017aa,Turchet:2017ab}.

The generated music as it is now does never use notes outside the ones in the selected chords~\cite{Mandanici:2014aa}. An added layer could generate more advanced melodies that are not so strongly restricted by the underlying chord. The chosen melodies should be able to adapt to the changes in the input, possibly defining more hierarchical levels of different note densities~\cite{simonetta_symbolic_2018}. The system should also embed an expressive performance generation, that should use the same input information to vary the produced melody~\cite{carnovalini_multilayered_2019, Cristani:2010aa}. 

From the gaming point of view, the system could benefit form a simpler and more straight-forward interface, and possibly from visual animations to accompany the music generation. 
Finally, a single player version of the game could be considered. Since this work was based on the interaction between players, a single player version would not be very reasonable. Yet, some of the categories of users that could benefit from this system, like motor rehabilitation patients, do not need the social aspect of this game to obtain the desired benefits. Moreover, a single player version could be enjoyed by players who simply want a game of musical interaction, regardless of the therapeutic benefits it could bring.

%
\vspace{-2mm}
\begin{acks}
This work is funded by a doctoral grant by University of Padova.
\end{acks}
\vspace{-2mm}
%
\bibliographystyle{ACM-Reference-Format}
\bibliography{biblio,other}

\end{document}